\documentclass{jfm}

\usepackage{graphicx}
\usepackage{epsfig}
\usepackage{newtxtext}
\usepackage{newtxmath}
\usepackage{natbib}
\usepackage{hyperref}
\usepackage[caption=false]{subfig}
\hypersetup{
    colorlinks = true,
    urlcolor   = blue,
    citecolor  = black,
}

\newcommand{\RomanNumeralCaps}[1]
\linenumbers
\newcommand{\St}{{\textit{St}}}
\newcommand{\Stbar}{{\bar{\textit{St}}}}

\newcommand{\bu}{\boldsymbol{u}}
\newcommand{\Xp}{\ensuremath{\boldsymbol{X}}}
\newcommand{\Vp}{\ensuremath{\boldsymbol{V}}}
\newcommand{\Op}{\ensuremath{\boldsymbol{\Omega}}}
\newcommand{\be}{\ensuremath{\boldsymbol{e}}}

\newcommand{\dd}{{\ensuremath{\mathrm{d}}}}

\title{Velocity and acceleration statistics of heavy spheroidal particles in turbulence}

\author{Sofia Allende\aff{1}
  and Jérémie Bec\aff{2,3}}

\affiliation{\aff{1}Earth and Climate Research, Earth and Life Institute, Université catholique de Louvain, Louvain-la-Neuve, Belgium
\aff{2}Universit\'e C\^ote d'Azur, Inria, CNRS, Calisto team, Sophia Antipolis, France
\aff{3}Mines Paris, PSL University, CNRS, Cemef, Sophia Antipolis, France}

\begin{document}
\maketitle

\begin{abstract}
Non-spherical particles transported by turbulent flow have a rich dynamics that combines their translational and rotational motions. Here, the focus is on small, heavy, inertial particles with a spheroidal shape fully prescribed by their aspect ratio. Such particles undergo an anisotropic, orientation-dependent viscous drag with the carrier fluid flow whose associated torque is given by the Jeffery equations. Direct numerical simulations of homogeneous, isotropic turbulence are performed to study systematically how the translational motion of such spheroidal particles depends on their shape and size. Surprisingly, it is found that the Lagrangian statistics of both velocity and acceleration can be thoroughly described in terms of an effective Stokes number obtained as an isotropic average over angles of the particle's orientation.  Corrections to the translational motion of particles due to their non-sphericity and rotation can hence be fully recast as an effective radius obtained from such a mean.
\end{abstract}

\section{Introduction}
\label{sec:intro}
Small complex particles suspended in a turbulent flow occur in a wide variety of natural processes. They are present in the oceans as phytoplankton \citep{sengupta2017phytoplankton} and in the atmosphere as volcanic ashes \citep{del2015experimental} or sea-salt aerosols \citep{grythe2014review}. These instances play a key role in climate balances: the atmospheric concentration of CO$_2$ is partly regulated by phytoplankton blooms \citep{leblanc2018nanoplanktonic}, while the Earth's radiative budget is largely impacted by airborne particles such as ashes and salts \citep{prata2019passive,horowitz2020effects}. The dynamics of such particles involve intricate internal and external physical interactions, the proper characterisation of which is still a challenge. 

We are here interested in understanding and modelling the join effects of particles inertia and non-sphericity on their transport by a turbulent flow and for that, we focus on small, heavy, ellipsoidal particles. In the case of spherical particles, the influence of inertia has been intensively studied, both numerically and experimentally~\citep[see, for instance,][and references therein]{brandt2022particle}. In particular, it is now well known that inertia has two important signatures:  \textit{preferential sampling}, whereby particles are ejected from rotation-dominated regions of the flow and concentrate in those with high strain; and \textit{filtering}, which is due to the delays that the particles have in following the fluid~\citep{toschi2009lagrangian}. The first effect dominates at low inertia and the particles oversample the most energetic regions of the flow, leading to velocity fluctuations slightly higher than the fluid. The second effect takes precedence at larger inertia, leading to a noticeable reduction of the particle kinetic energy. Concerning particle acceleration, these two mechanisms contribute both to a significant depletion of the most violent fluctuations. However, much less is known about non-spherical particles, for which translational and rotational dynamics are a priori tightly coupled. For example, preferential sampling could be significantly modified by the \textit{preferential alignment} of the particles with the local geometrical structure of the flow.

The simplest instance of non-spherical particles is axisymmetric ellipsoids, also referred to as \textit{spheroids}, whose shape is defined by a single parameter, their aspect ratio. Since \citet{jeffery1922motion}, explicit equations for their translational and rotational motion have been known in the limit where their Reynolds number is zero. In recent years, significant work has been devoted to the dynamics of spheroidal particles in turbulent flows \citep[see][]{voth2017anisotropic}. Two main issues have received attention. The first concerns the rotation rate of particles and how the relative contributions of spinning and tumbling depend on their shape. In the absence of inertia, rod-like particles in homogeneous isotropic flow align their axis of symmetry with the fluid vorticity, while disk-shaped particles have it orthogonal~\citep{pumir2011orientation,ni2014alignment}. As a result of this preferential alignment, inertialess prolate particles have a higher spinning rate than oblate ones, and vice versa for the tumbling rate \citep{parsa2012rotation,marcus2014measurements,byron2015shape}, with similar observations in inhomogeneous, anisotropic channel flows~\citep{marchioli2013rotation,baker2022experimental}. Inertia reduces both the tumbling and spinning rates because of preferential sampling~\citep{gustavsson2014tumbling,zhao2015rotation,roy2018inertial}.  The second issue studied at length is the effect of non-sphericity on the gravitational settling of particles. Small heavy spheroids tend to fall with a preferential orientation that fluctuates under the action of turbulence~\citep{klett1995orientation,siewert2014orientation,anand2020orientation}, with significant effects on their collision rates~\citep{siewert2014collision,jucha2018settling}. However, this problem is rather delicate, because the inertial torque of the fluid plays a dominant role~\citep{gustavsson2019effect,sheikh2020importance}.

Previous studies on spheroidal particles have hence primarily focused on their orientation dynamics, with less attention given to their translational motion. However, \citet{shapiro1993deposition} and \citet{zhang2001ellipsoidal} suggested that shape effects on the deposition velocity of spheroids can be cast as an effective Stokes number based on an isotropic average of the particle mobility tensor (inverse of its drag/resistance).  Direct numerical simulations by \citet{mortensen2008dynamics} and \citet{challabotla2015orientation} in turbulent channel flow confirm that average translational motions weakly depend on the aspect ratio for spheroids with the same effective Stokes number. In this work, we focus on the fine turbulent fluctuations of the velocity and acceleration of heavy spheroids transported by a homogeneous isotropic flow. We provide strong evidence that these statistics, including rare events, are fully characterised in terms of the effective, isotropically-averaged Stokes number. Such findings imply that the translational dynamics of small inertial spheroids is undistinguishable from that of small heavy spheres with an equivalent shape-dependent radius.

This paper is structured as follows:  In \S\ref{sec:dynamics}, we briefly overview the equations of motion for inertial spheroids, discuss associated timescales, and present our numerical simulations. In \S\ref{sec:results}, we report and discuss our main results on the statistics of particle translational velocities and accelerations. Finally, in \S\ref{sec:conclusions}, we draw conclusions and offer perspectives for future work.

\section{Dynamics of small inertial spheroids, timescales, and numerical methods}
\label{sec:dynamics}
\begin{figure}
 \centerline{
  \subfloat{\includegraphics[width=0.32\linewidth]{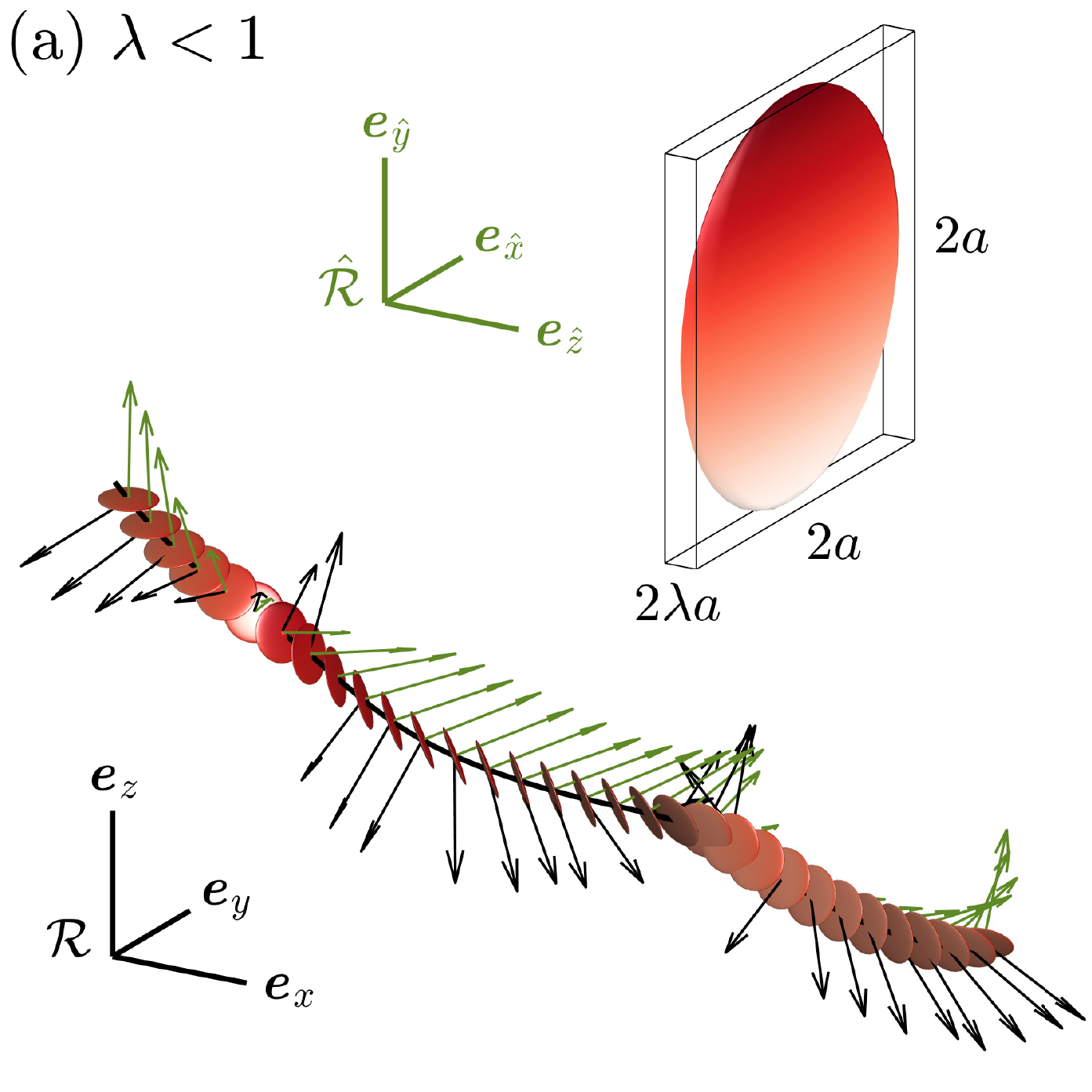}\label{fig:oblate}} \
  \subfloat{\includegraphics[width=0.32\linewidth]{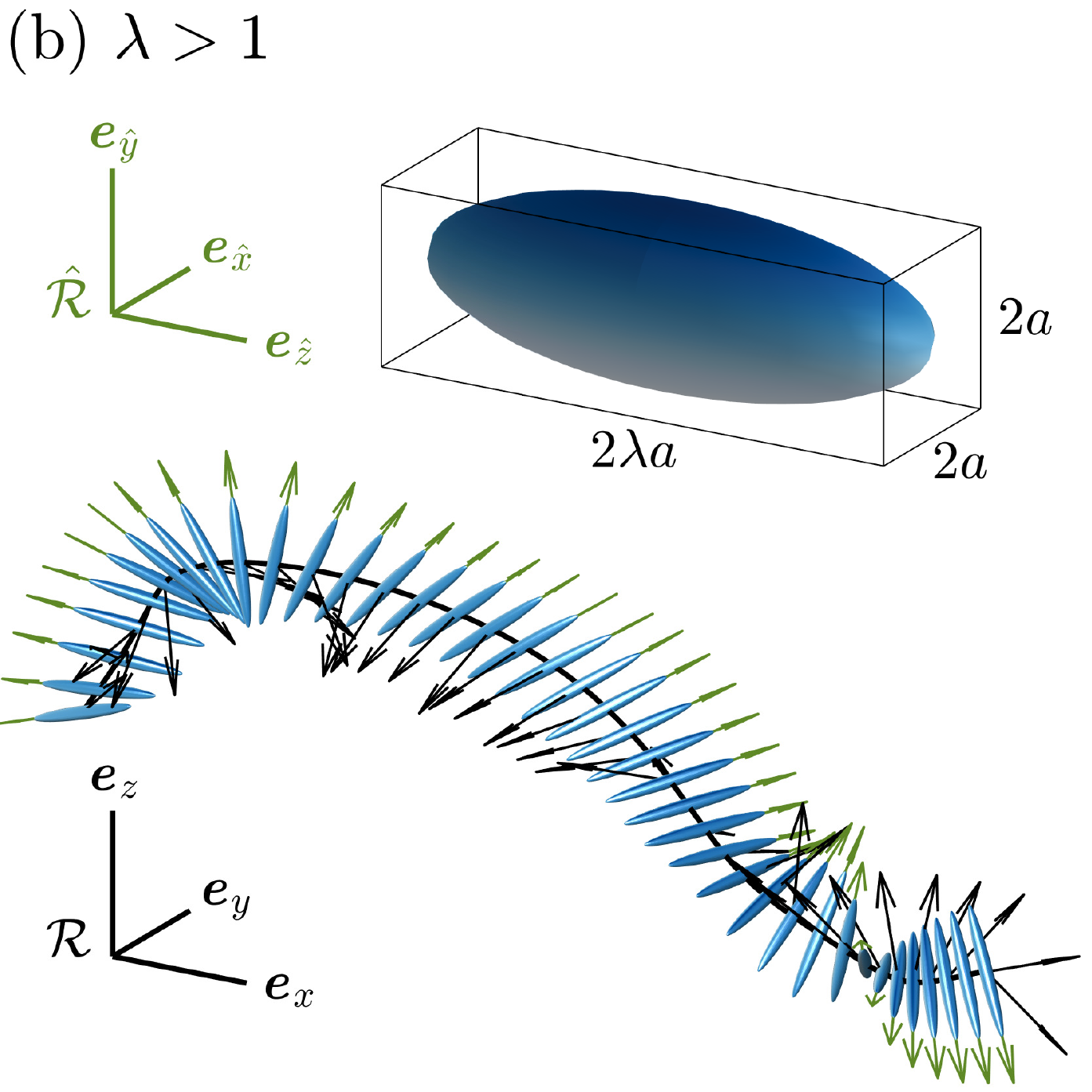}\label{fig:prolate}} \
  \subfloat{\includegraphics[width=0.32\linewidth]{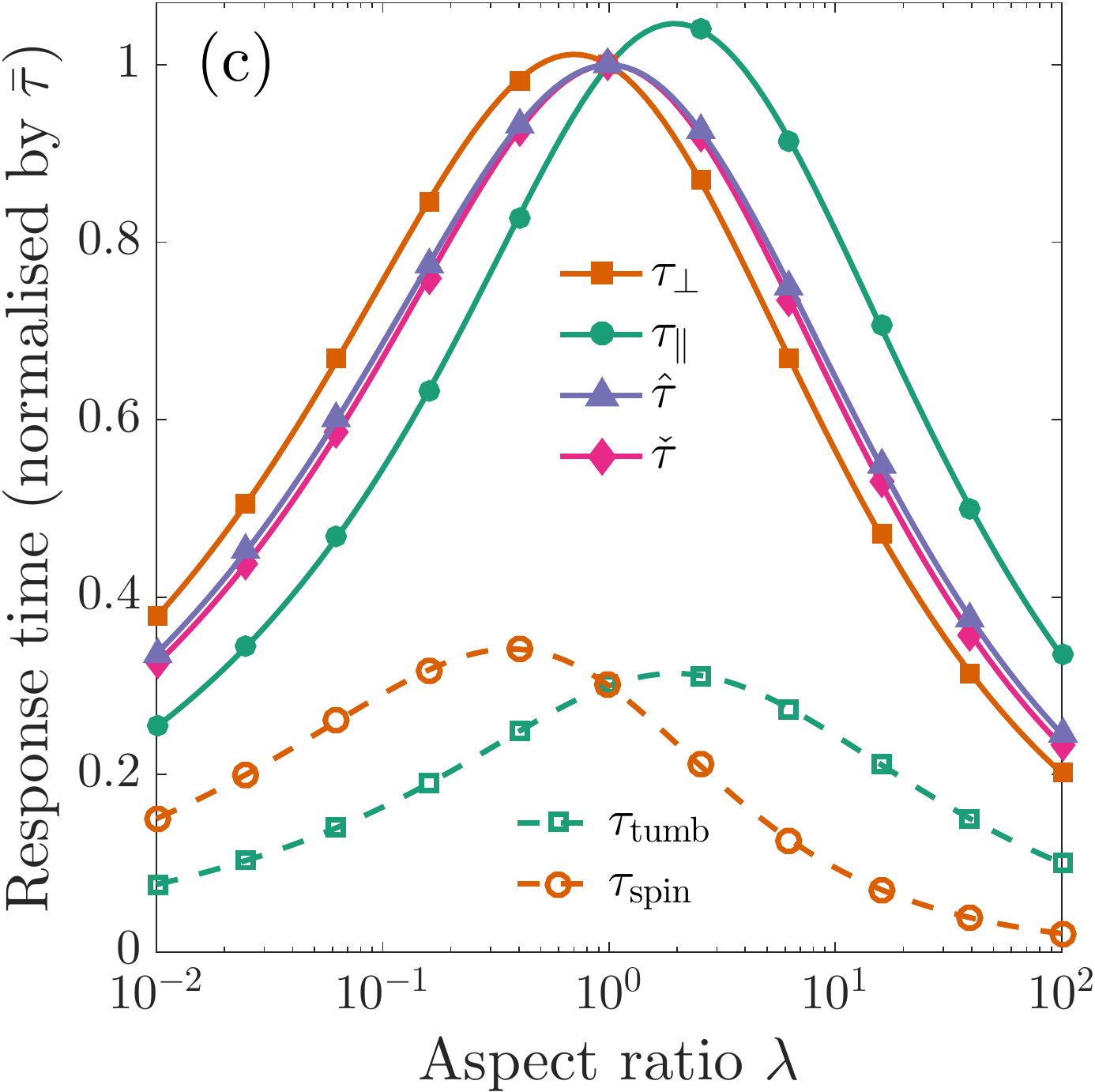}\label{fig:timescales}}}
\caption{(a) Oblate particle, together with a representative trajectory on which green arrows stand for particle orientation $\boldsymbol{e}_{\!\hat{z}}$ and black ones for its acceleration $\dd\Vp/\dd t$. (b) Same for a prolate particle. (c) Various response times (see text for definitions), normalised by the equal-mass time  $\bar{\tau}=2\rho_\mathrm{p}\lambda^{2/3} a^2 /(9\rho_\mathrm{f} \nu)$, as a function of the particle aspect ratio.}
\end{figure}

We focus on spheroidal particles, which are ellipsoids of revolution with two equal semi-axes $a=b$ and a principal axis $c = \lambda\,a$. The aspect ratio $\lambda$ characterises their shape: Oblate particles have $\lambda < 1$, spheres $\lambda = 1$, and prolate particles $\lambda >1$ (see Fig.~\ref{fig:oblate} and \ref{fig:prolate}).
We consider the case where such particles are suspended in a developed, incompressible turbulent velocity field $\bu(\boldsymbol{x},t)$ and are much smaller than the associated Kolmogorov dissipative scale, namely $a,b,c \ll \eta = \nu^{3/4}/\varepsilon^{1/4}$, where $\nu$ is the fluid kinematic viscosity and $\varepsilon$ the mean dissipation rate of kinetic energy. We moreover assume that the particles are much heavier than the surrounding fluid, \textit{i.e.}\/ their mass density $\rho_{\rm p}$ is much larger than the fluid density $\rho_{\rm f}$, and that their velocity relative to the fluid, together with their size, defines an infinitely small Reynolds number. Finally, the particles are presumed sufficiently dilute to neglect their feedback on the flow and turbulence sufficiently intense to ignore the effects of gravity.

The dynamics of non-spherical particles involves both translational and rotational motions. The translation is determined by the position $\Xp$ and velocity $\Vp$ of the particle's center of mass in the inertial frame of reference $\mathcal{R} = (\be_{\!x}, \be_{\!y}, \be_{\!z})$ of the fluid flow. Under the above assumptions, they are given by the linear momentum equations \citep{brenner1964stokes}
\begin{equation}
  \frac{\dd\Xp}{\dd t} = \Vp, \qquad \frac{\dd\Vp}{\dd t} = -\frac{1}{\bar{\tau}}\,\mathsfbi{A}^{\mathsf{T}}\mathsfbi{D}\,\mathsfbi{A}\, \left[\Vp - \bu(\Xp,t)\right],
    \label{eq:trans_spheroids}
\end{equation}
where we introduced the response time $\bar{\tau}=2\rho_\mathrm{p}\lambda^{2/3} a^2 /(9\rho_\mathrm{f} \nu)$ associated to a spherical particle with the same mass as the spheroid. $\mathsfbi{A}\,$ denotes the rotation matrix that maps $\mathcal{R}$ to the reference frame  $\hat{\mathcal{R}} = ({\be}_{\!\hat{x}}, {\be}_{\!\hat{y}}, {\be}_{\!\hat{z}})$ of the particle, ${\be}_{\!\hat{z}}$ being along its revolution axis (see Fig.~\ref{fig:oblate} and~\ref{fig:prolate}). The drag tensor is expressed in $\hat{\mathcal{R}}$ where it is diagonal, $\mathsfbi{D}= \mathrm{diag}[\mathsfi{D}_\perp, \mathsfi{D}_\perp, \mathsfi{D}_\parallel]$, with 
\begin{equation}
\mathsfi{D}_\perp\!= \frac{8 (\lambda ^2\!-\!1)}{3\lambda^{1/3}\!\left[\chi(2\lambda^2\!-\!3)\!+\!\lambda\right]},\ \ \ \mathsfi{D}_\parallel\!=\frac{4 (\lambda ^2\!-\!1)}{3\lambda^{1/3}\!\left[\chi(2 \lambda ^2\!-\!1)\!-\!\lambda\right]},\ \ \ \chi\!=  \frac{\log\!\left(\lambda\!+\!\sqrt{\lambda^2\!-\!1}\right)}{\sqrt{\lambda^2\!-\!1}}.
\label{eq:drag_coeffs}
\end{equation}
Note that  when $\lambda<1$, the factor $\chi$ involves the logarithm of a unit-modulus complex number and  reduces to $\chi\!=\!\arctan(\sqrt{1\!-\!\lambda^2}/\lambda)/\sqrt{1\!-\!\lambda^2}$, as usually stipulated for oblate particles.

The  components of $\mathsfbi{D}$ define two shape-dependent time scales, $\tau_\perp\!=\!\bar{\tau}/\mathsfi{D}_\perp$ and $\tau_\parallel\!=\!\bar{\tau}/\mathsfi{D}_\parallel$ associated to the drag in the directions perpendicular and parallel to the particle axis of symmetry, respectively. \citet{shapiro1993deposition} introduced an effective response time $\hat{\tau}$ by performing an isotropic average of the mobility tensor $\bar{\tau} \mathsfbi{D}^{-1}$ over all orientations:
\begin{equation}
\hat{\tau} = \tfrac{1}{3}\mathrm{tr}\big(\mathsfbi{D}^{-1}\big)\,\bar{\tau}= \frac{2\tau_\perp\!+\tau_\parallel}{3} = \bar{\tau}\,\lambda^{1/3}\chi = \frac{2\rho_\mathrm{p}\,\lambda\,a^2}{9\rho_\mathrm{f}\,\nu}\,\frac{\log\!\left(\lambda+\!\sqrt{\lambda^2\!-\!1}\right)}{\sqrt{\lambda^2\!-\!1}}.
\label{eq:deftauhat}
\end{equation}
\citet{fan1995dispersion} considered an alternative effective response time $\check{\tau}$ obtained  by averaging the drag tensor rather than the mobility matrix. This corresponds to the harmonic mean of the orientation-dependent response times, namely
\begin{equation}
\check{\tau} = \left(\tfrac{1}{3}\mathrm{tr}\big(\mathsfbi{D}\,\big)\right)^{\!-1}\!\bar{\tau}=\left(\frac{2\tau_\perp^{-1}\!+\tau_\parallel^{-1}}{3}\right)^{\!\!-1}\!\! = 
\frac{\rho_\mathrm{p}\,\lambda\,a^2}{2\rho_\mathrm{f}\,\nu}\,\frac{4\chi^2(\lambda^2\!-\!1)^2 - (\lambda\!-\!\chi)^2}{\left(\lambda ^2\!-\!1\right) \left[10 \chi(\lambda^2\!-\!1)  -3(\lambda\!-\!\chi)\right]}.
\label{eq:deftaucheck}
\end{equation}
The dependence of the two response times $\hat{\tau}$ and $\check{\tau}$ upon the aspect ratio $\lambda$ is shown in Fig.~\ref{fig:timescales}. Surprisingly, their difference hardly exceeds a couple of percents. One finds that $\Delta\!=\!(\hat{\tau }-\check{\tau })/\bar{\tau}$ behaves as $\Delta\!\simeq\!(2/225)(\lambda-1)^2$ for nearly spherical particles ($\lambda\!\simeq\!1$), as $\Delta\!\simeq\!(\pi/56)\lambda^{1/3}\!$ for thin disks ($\lambda\!\ll\!1$), and as $\Delta\!\simeq\! (1/10)(\log \lambda)\lambda^{-2/3}\!$ for slender fibers ($\lambda\!\gg\!1$). Such tiny differences make it almost impossible to select the most relevant of these two response times from numerical or experimental data.

Turning now to rotational dynamics, the orientation matrix $\mathsfbi{A}$ evolves with an angular velocity $\Op$, which is more conveniently expressed in the particle reference frame $\hat{\mathcal{R}}$, so that
\begin{equation}
  \frac{\dd \mathsfbi{A}}{\dd t} = -\begin{bmatrix} \,0 &-\Omega_{\hat{z}} & \phantom{-}\Omega_{\hat{y}}\\ \phantom{-}\Omega_{\hat{z}} & \,0 &-\Omega_{\hat{x}}\\ -\Omega_{\hat{y}}& \phantom{-}\Omega_{\hat{x}} & \,0 \end{bmatrix}\,\mathsfbi{A}, \qquad \frac{\dd (\mathsfbi{I} \, \Op)}{\dd t} +
  \Op\times(\mathsfbi{I}\, \Op)= \boldsymbol{T},
\label{eq:rot_spheroids}
\end{equation}
where $\mathsfbi{I}= (4/15)\upi\,\rho_{\rm p}\,\lambda\,a^5\,\mathrm{diag}[1\!+\!\lambda^2,1\!+\!\lambda^2,2]$ is the spheroid's moment of inertia  about its principal axes and $\boldsymbol{T}$ the hydrodynamic torque acting on the particle. It reads \citep{jeffery1922motion}
\begin{equation}
  \boldsymbol{T} = \left[
  \begin{array}{l} T_{\hat{x}}\\ T_{\hat{y}}\\ T_{\hat{z}}  \end{array}\right]
  =\frac{16}{3}\upi\rho_{\rm f}\,\nu\,a^3 \left[
  \begin{array}{r}
  -\frac{\lambda^4-1}{\chi(2 \lambda^2-1)-\lambda}\left(\Omega_{\hat{x}}-\tfrac{1}{2}\omega_{\hat{x}}+\frac{\lambda ^2-1}{\lambda^2+1}\mathsfi{S}_{\hat{z}\hat{y}}\right)\\[5pt]
    -\frac{\lambda^4-1}{\chi(2 \lambda^2-1)-\lambda}\left(\Omega_{\hat{y}}-\tfrac{1}{2}\omega_{\hat{y}}-\frac{\lambda ^2-1}{\lambda^2+1}\mathsfi{S}_{\hat{x}\hat{z}}\right)\\[5pt]
   -\frac{\lambda ^2-1}{\lambda -\chi}\left(\Omega_{\hat{z}}-\tfrac{1}{2}\omega_{\hat{z}}\right)
  \end{array}\right]\!,
  \label{eq:jefferytorque}
\end{equation}
and depends on the fluid vorticity $\boldsymbol{\omega} = \boldsymbol{\nabla}\times\bu$ and strain tensor $\mathsfbi{S} =(\boldsymbol{\nabla} \bu +\boldsymbol{\nabla} \bu^{\!\mathsf{T}})/2$, both evaluated at the particle position and in the rotating reference frame $\hat{\mathcal{R}}$.

Equations~(\ref{eq:rot_spheroids}) and (\ref{eq:jefferytorque}) define two rotational response times associated to the particle tumbling (rotation along the $\hat{x}$ and $\hat{y}$ axes) and to its spinning (rotation along the spheroid's axis of symmetry $\hat{z}$). They read
\begin{equation}
	\tau_\mathrm{tumb} = \frac{\rho_{\rm p}\,\lambda\,a^2}{20\rho_{\rm f}\,\nu}\,\frac{\chi(2 \lambda^2\!-\!1)\!-\!\lambda}{\lambda ^2\!-\!1}= \frac{3}{10}\tau_\parallel,\quad \tau_\mathrm{spin} = \frac{\rho_{\rm p}\,\lambda\,a^2}{10\rho_{\rm f}\,\nu}\,\frac{\lambda\!-\!\chi}{\lambda^2\!-\!1} =\frac{3}{10}(2\tau_\perp\!-\!\tau_\parallel).
\end{equation}
These two times are displayed in Fig.~\ref{fig:timescales}. As stressed by \citet{zhao2015rotation} and \citet{marchioli2016relative}, they remain shorter than the translational timescales for all values of the aspect ratio. More precisely, one can actually show that both the tumbling and the spinning response times are smaller than $(9/20)\hat{\tau}$ for all $\lambda>0$, this bound being attained by $\tau_\mathrm{tumb}$ for $\lambda\to\infty$ and by $\tau_\mathrm{spin}$ for $\lambda\to0$. The separation of timescales between the translational and rotational motions of the particle is hence clear, but one can question whether a factor $\approx\!2$ is enough to assume complete decoupling. To address this, we have recourse to direct numerical simulations (DNS) and investigate shape dependence in the statistics of heavy spheroids transported by a homogenous isotropic turbulent flow.

The three-dimensional incompressible Navier--Stokes equations with large-scale forcing are integrated using the parallel pseudo-spectral solver \emph{LaTu} with third-order Runge--Kutta time marching~\citep[see][]{homann2007impact}. Relevant simulation parameters are summarised in Tab.~\ref{tab:num_1024}. Equations (\ref{eq:trans_spheroids}) and (\ref{eq:rot_spheroids}) for the dynamics of spheroidal particles are integrated numerically with the same time stepping as the fluid flow. The orientation of each individual particle is represented as a quaternion \citep[see][]{mortensen2008dynamics,siewert2014orientation} to ease numerical integration and stability. In order to span the particles parameter space, we have considered $90$ different families of $500\,000$ particles each. They combine nine different aspect ratios, ranging from $\lambda = 0.1$ to $10$, and ten response times, spanning $\Stbar= \bar{\tau}/\tau_\eta = 0.1$ to $6.4$, where $\tau_\eta=(\nu/\varepsilon)^{1/2}$ denotes the Kolmogorov dissipative time scale. The properties of spheroids are stored with a period $\approx 6.8 \,\tau_\eta$. Statistics are computed over approximately $4$ large-eddy turnover times after a statistical steady state is reached.

\begin{table}
\begin{center}
\begin{tabular}{c c c c c c c c c c }
  $N^3$ & $\nu$ & $\Delta t$ & $\varepsilon$  & $\eta$
  & $\tau_\eta$ & $u_\mathrm{rms}$  & $L$ & $\tau_L$ & $\Rey_\lambda $\\
  $1024^3$ & $2\times10^{-4}$ & $5\times10^{-4}$ & $0.431$ & $0.0021$
  & $0.022$ &  $0.87$ &  $1.52$ & $1.753$ & $315$ \\
  \end{tabular}
  \caption{\textbf{DNS parameters.} $N^3$ number of collocation points, $\nu$ kinematic viscosity, $\Delta t$ time step, $\varepsilon$ average dissipation rate, $\eta = (\nu^{3}/\varepsilon)^{1/4}$ Kolmogorov dissipative scale, $\tau_\eta = (\nu/\varepsilon)^{1/2}$ Kolmogorov time, $u_\mathrm{rms}$ root-mean square velocity, $L = u_\mathrm{rms}^3/\varepsilon$ large scale, $\tau_L = L/ u_\mathrm{rms}$ large-eddy turnover time, $\Rey_\lambda = \sqrt{15}\, u_\mathrm{rms}^2/(\nu\varepsilon)^{1/2}$ Taylor-scale Reynolds number.}
  \label{tab:num_1024}
  \end{center}
\end{table}

\section{Results and discussions}
\label{sec:results}

\subsection{Fluctuations of particle velocities}
\begin{figure}
  \centering
    \subfloat{\includegraphics[width=0.49\linewidth]{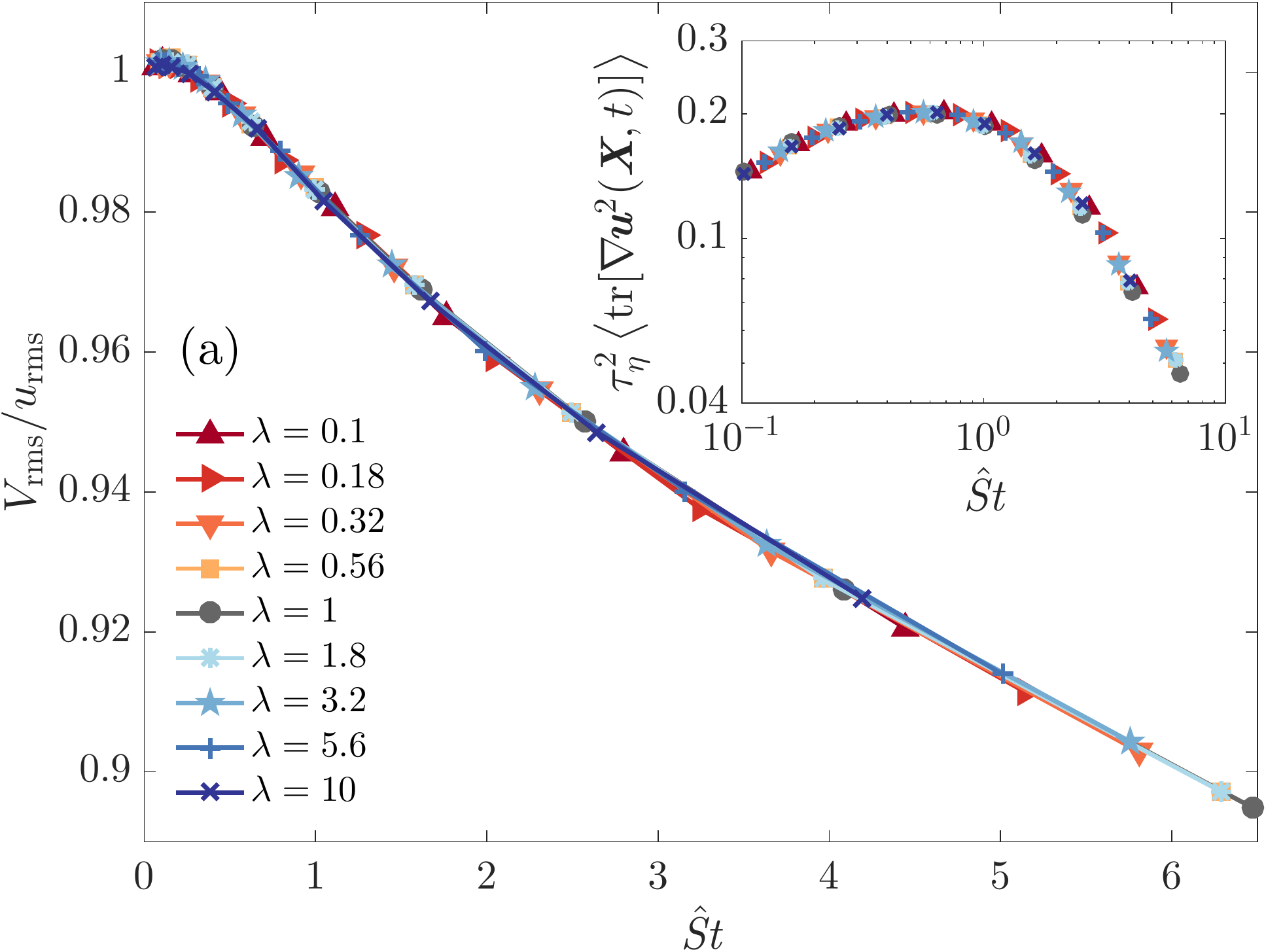}\label{fig:vrms}}
    \hfill
    \subfloat{\includegraphics[width=0.49\linewidth]{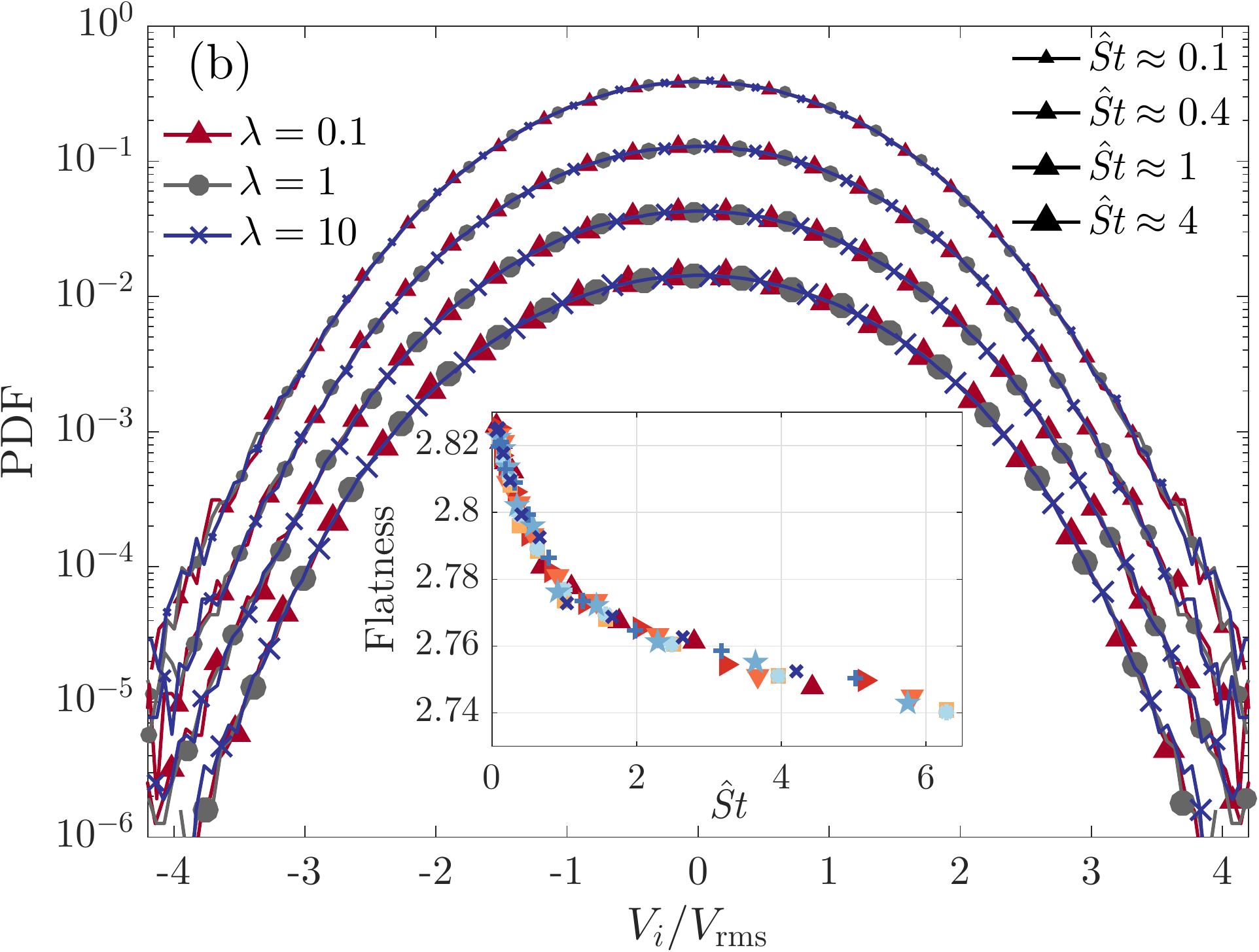}\label{fig:pdfv}}
    \caption{(a) Root-mean-squared particle velocity $V_{\rm rms}$ as a function of the isotropically-averaged Stokes number $\hat{\St}=\hat{\tau}/\tau_\eta$. \textit{Inset:} Average of $\mathrm{tr}(\nabla\bu)^2$ along particle paths versus $\hat{\St}$. (b)~Probability density functions (PDF) of the particle velocity components (normalised to unit variance) for combinations of $\bar{\tau}$ and $\lambda\in\{0.1,1,10\}$ associated to four different values of $\hat{\St}$ (as labeled); Data has been shifted to increase visibility. \textit{Inset:} Flatness $\langle V_i^4\rangle / V_{\rm rms}^4$ of the particle velocity as a function of $\hat{\St}$ for the various aspect ratios, labelled as in panel~a.}
\end{figure}

To analyse the translational motion of inertial spheroidal particles, we start by measuring the fluctuations in the velocity of their center of mass along Lagrangian trajectories. For this purpose, we introduce the particle root-mean-squared (rms) velocity
\begin{equation}
  V_{\rm rms} = \left\langle V_i^2 \right\rangle^{1/2} = \left\langle \tfrac{1}{3}|\Vp|^2\right\rangle^{1/2},
\end{equation}
where we recall that $\Vp = (V_x, \, V_y, \, V_z)$ is the particle translational velocity in the reference frame of the fluid flow and  $\langle \cdot\rangle$ denotes averages over both time and particle initial positions. Figure~\ref{fig:vrms} shows the particle rms velocity as a function of the isotropically-averaged Stokes number $\hat{St} = \hat{\tau}/\tau_\eta$, which is obtained by non-dimensionalising the response time $\hat{\tau}$ defined in Eq.~(\ref{eq:deftauhat}) by the Kolmogorov timescale. We observe that the data collected across a range of aspect ratios $\lambda$ collapse on the top of each other when plotted as a function of $\hat{St}$. This indicates that the variance of the particle velocity is effectively as if given by an average over all its possible orientations. Note that, as argued in the previous section, a similar collapse in the data is observed if, instead of $\hat{\St}$, we use the harmonically-averaged Stokes number $\check{\St} = \check{\tau}/\tau_\eta$, with $\check{\tau}$ given by (\ref{eq:deftaucheck}). Interestingly, our data indicate that the approach to the tracer limit $\hat{\St}\to0$, where particle rms velocities converge to that of the fluid $u_{\rm rms}$, is independent of particle shape up to statistical errors. In addition, the depletion of particle velocities occurring at large values of $\hat{\St}$, which is generally attributed to filtering of the fluid velocity, also seems independent of the aspect ratio $\lambda$. Finally, we find that at intermediate Stokes numbers, $\hat{\St}\lesssim 0.3$, particles rms velocities are slightly larger than $u_{\rm rms}$. This feature has previously been observed for spherical particles \citep{salazar2012inertial_vel} and is due to the preferential sampling of energy-containing, strain-dominated regions of the flow by heavy inertial particles.

Preferential sampling is generally quantified by evaluating the mean trace of the squared fluid-velocity gradient $\left\langle\mathrm{tr}[\nabla\bu^2(\Xp,t)]\right\rangle$ along particle paths. The inset of Fig.~\ref{fig:vrms} shows this quantity for various aspect ratios $\lambda$. Once again, the measurements collapse on the top of each other when plotted as a function of the isotropically-averaged Stokes number $\hat{St}$. This confirms that the relevant response time of the spheroids is given by $\hat{\tau}$ (or $\check{\tau}$, up to statistical precision), rather than the time $\bar{\tau}$ associated to a spherical particle with equivalent mass. 

Higher-order velocity statistics also exhibit independence upon shape. Figure~\ref{fig:pdfv} shows the probability distributions of particle-velocity components for $\lambda=0.1$, $1$, and $10$, using selected values of the response time such that $\hat{\St}\approx 0.1$, $0.4$, $1$, and $4$. The distributions associated with different aspect ratios collapse onto master curves that depend solely upon $\hat{\St}$, indicating that the isotropically-averaged Stokes number fully accounts for particle shape at the level of one-time velocity fluctuations. This observation is further supported by the flatness of the particle-velocity distributions shown in the inset of Fig.~\ref{fig:pdfv}. Overall, our findings indicate that $\hat{\St}$ is a robust and informative parameter for characterizing the translational dynamics of heavy spheroidal particles in homogeneous isotropic turbulent flow.  

\subsection{Fluctuations of particle accelerations}
\begin{figure}
  \centering
  \subfloat{\includegraphics[width=0.49\linewidth]{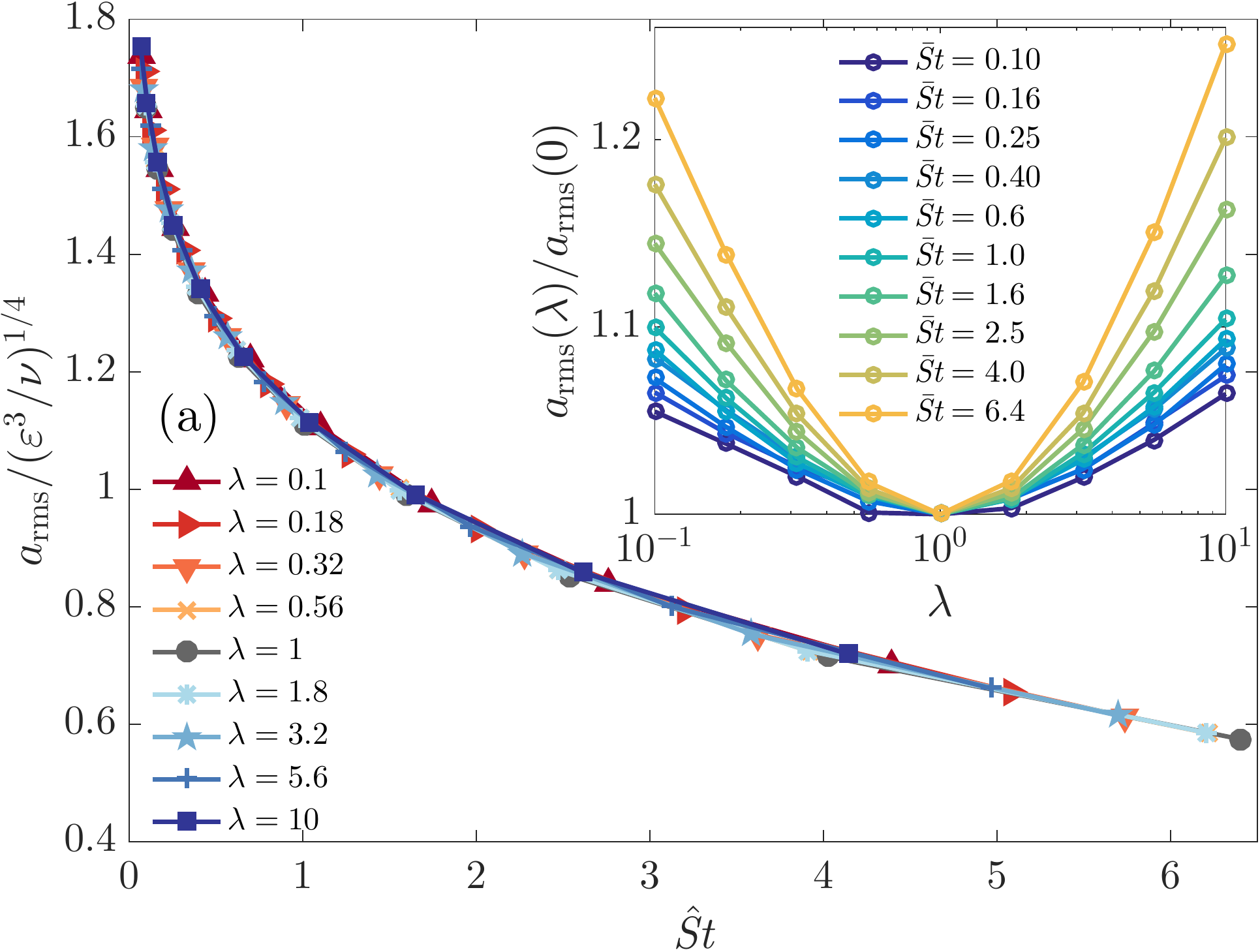}\label{fig:arms}}
  \hfill
  \subfloat{\includegraphics[width=0.49\linewidth]{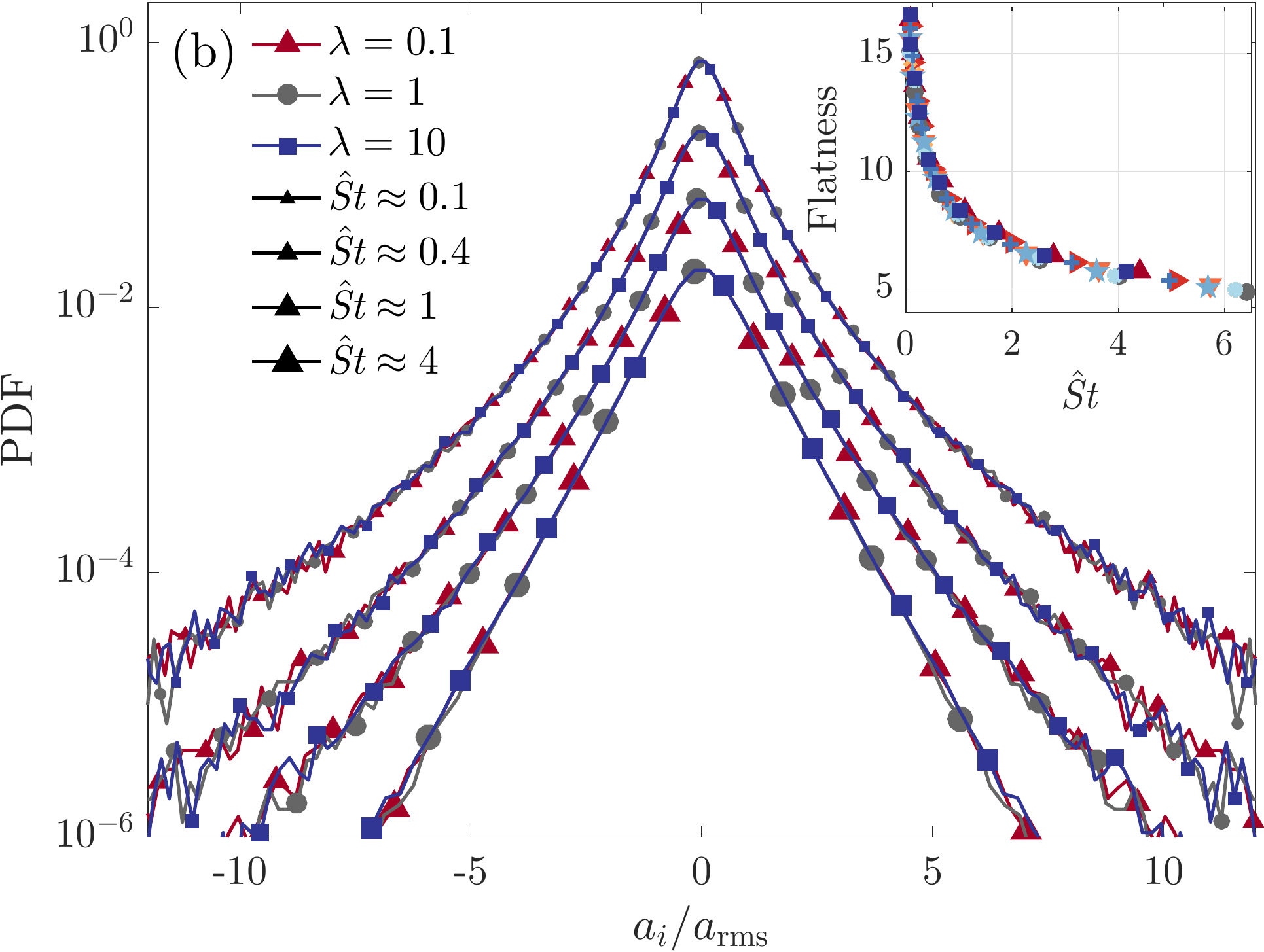}\label{fig:pdfa}}
  \caption{(a) Root-mean-square particle acceleration $a_{\rm rms}$ normalised to turbulent viscous units, as a function of $\hat{\St}$. \textit{Inset:} Same, as a function of $\lambda$ for various $\bar{\St}$, normalised here to the root-mean-square acceleration of a spherical particle with same mass. (b) PDF of the components of particle acceleration (normalised to unit variance) for combinations of $\bar{\tau}$ and $\lambda\in\{0.1,1,10\}$ associated to four different $\hat{\St}$ (as labeled); The curves have been shifted vertically to increase visibility. \textit{Inset:} Flatness $\langle ({\rm d}V_i/{\rm d}t)^4\rangle / a_{\rm rms}^4$ of the particle acceleration as a function of $\hat{\St}$ for the various aspect ratios, labelled as in~panel a.}
\end{figure}

In the previous subsection, it was observed that the orientation of inertial spheroids appears to be uncorrelated with their translational motion. However, this behaviour might be due to the fact that velocity fluctuations only involve large-scale turbulent eddies, and on such scales, the orientation may be effectively averaged in an isotropic manner. To investigate this possibility further, it is necessary to examine small-scale quantities, which motivates the study of particle accelerations. To do so, we begin by computing the particle root-mean-squared acceleration
\begin{equation}
  a_{\rm rms} = \langle a_i^2 \rangle^{1/2} = \langle |\boldsymbol{a}|^2/3 \rangle^{1/2},\quad\mbox{with } \boldsymbol{a} = \mathrm{d}\Vp/\mathrm{d}t.
\end{equation}
The inset of Fig.~\ref{fig:arms} illustrates the dependence of this quantity upon the particle aspect ratio $\lambda$. We observe that $a_{\rm rms}(\lambda)$ increases as non-sphericity becomes more pronounced, for all values of the equivalent-mass Stokes number $\bar{\St} = \bar{\tau}/\tau_\eta$. However, when the particle shape is incorporated into the Stokes number, the behaviour of a sphere is recovered, as in the case of velocity fluctuations. Figure~\ref{fig:arms} indeed shows that the rms components of the particle acceleration depend solely on the isotropically-averaged Stokes number $\hat{\St}$. It is worth noting that accelerations are there normalised by the dimensional value expected for fluid elements, which is proportional to namely $\eta/\tau_\eta^2 = \varepsilon^{3/4}/\nu^{1/4}$. Our measurements suggest $a_{\rm rms}/(\varepsilon^{3/4}/\nu^{1/4}) \approx 1.8$ in the limit $\hat{\St}\to 0$. This value, which depends on the fluid-flow Reynolds number, is consistent with previously reported measurements at $R_\lambda \approx 300$~\citep[see, \textit{e.g.},][]{yeung2006acceleration}. At large values of $\hat{\St}$, a continuous depletion of particle acceleration is observed, similar to the case of spherical particles. This depletion is caused by a complex interplay between preferential sampling and filtering, as discussed in \cite{bec2005acceleration}.

Figure~\ref{fig:pdfa} shows the probability density functions of the acceleration components. Similar to the velocity components, distributions with the same value of $\hat{\St}$ but different aspect ratios $\lambda$ collapse on the top of each other, even at large fluctuations. This trend is confirmed by the flatness of the distribution of particle acceleration shown as a function of $\hat{\St}$ in the inset of Fig.~\ref{fig:pdfa}. We emphasise that particle acceleration statistics depend solely on the isotropically-averaged Stokes number $\hat{\St}$, just like velocity statistics, and can thus be straightforwardly deduced from those of spheres. This suggests again that there is a fundamental independence between the orientation of spheroidal particles and the dynamics of their center of mass.

\subsection{Two-time statistics}

\begin{figure}
  \centering
  \subfloat{\includegraphics[width=0.48\linewidth]{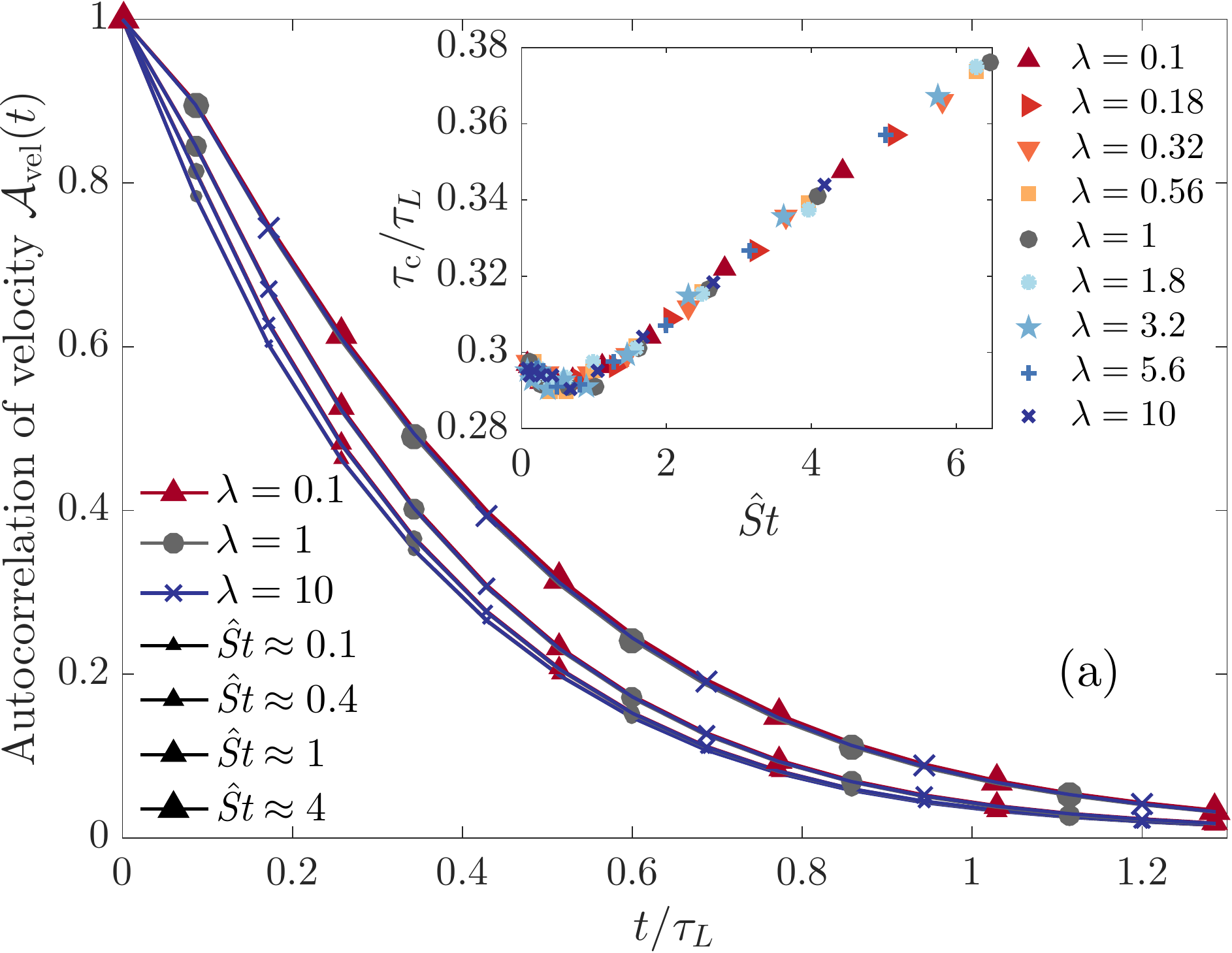}\label{fig:corrv}}
  \hfill
  \subfloat{\includegraphics[width=0.48\linewidth]{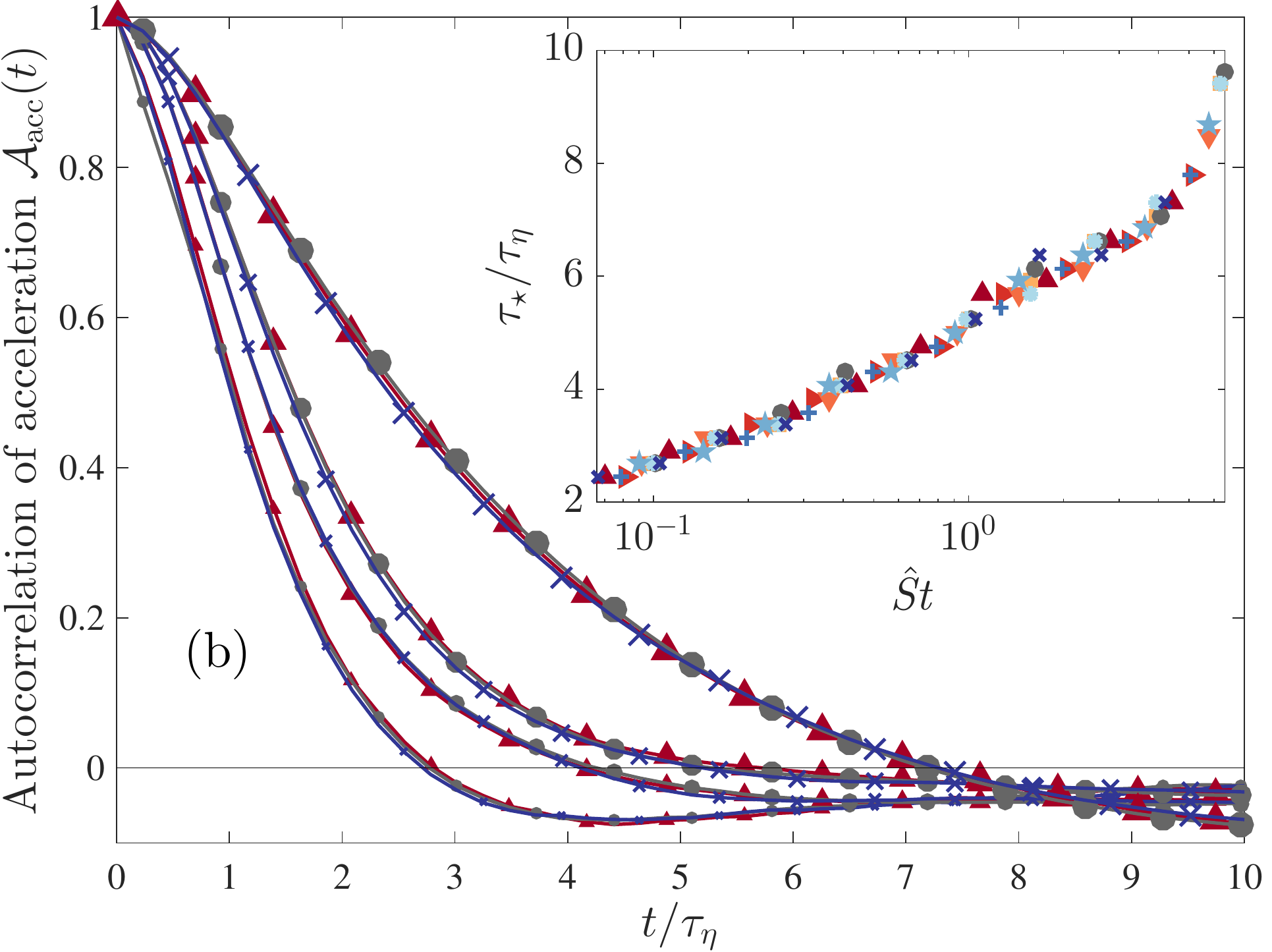}\label{fig:corra}}
  \caption{(a)~Time autocorrelations $\mathcal{A}_{\rm vel}$ of particle velocity components for different combinations of $\lambda$ and $\bar{\tau}$  associated to four values of $\hat{\St}$, as labelled. \textit{Inset:}~Correlation times $\tau_{\rm c}$ obtained from exponential fits of $\mathcal{A}_{\rm vel}$, plotted here as a function of $\hat{\St}$. (b)~Time autocorrelations $\mathcal{A}_{\rm acc}$ of particle acceleration for the same aspect ratios and isotropically-averaged Stokes numbers as in panel a. \textit{Inset:}~Zero-crossing time $\tau_\star$ of $\mathcal{A}_{\rm acc}$.}
\end{figure}

To highlight further the relevance of the isotropically-averaged Stokes number, we study the time autocorrelations of the particle velocity and acceleration components
$$\mathcal{A}_{\rm vel}(t) = \left\langle \Vp(t)\cdot\Vp(0) \right\rangle/(3V_{\rm rms}^2), \qquad \mathcal{A}_{\rm acc}(t) =
 \langle \boldsymbol{a} (t)\cdot \boldsymbol{a}(0) \rangle /(3a_{\rm rms}^2).$$
The time correlation of particle velocity is shown for selected values of $\hat{\St}$ and different aspect ratios $\lambda$ in Fig.~\ref{fig:corrv}. These results provide further evidence for the importance of the isotropically-averaged Stokes number.  The autocorrelation $\mathcal{A}_{\rm vel}(t)$ exhibits an exponential decay $\propto \exp(-t/\tau_{\rm c})$, allowing for an estimate of the Lagrangian correlation time $\tau_{\rm c}$ of particle velocity. In the inset of Fig.~\ref{fig:corrv}, we plot $\tau_{\rm c}$ as a function of $\hat{\St}$. For tracers, it is known that $\tau_{\rm c} \approx 0.3\,\tau_L$ \citep[see, \textit{e.g.},][]{yeung1989lagrangian}, and this is recovered in our simulations. At large inertia, $\tau_{\rm c}$ grows almost linearly as a function of $\hat{\St}$. Interestingly, for $\hat{\St}\lesssim1$,  the correlation time is slightly reduced by inertia, consistent with our previous observation that particles with small-to-intermediate inertia tend to oversample the highly energetic regions of the flow.

Figure~\ref{fig:corra} confirms the relevance of $\hat{\St}$ for two-time acceleration statistics. The correlation time can now be estimated by the zero-crossing time $\tau_\star$, defined as the smallest time at which $\mathcal{A}_{\rm acc}(\tau_\star)=0$.  The inset of Fig.~\ref{fig:corra} shows that $\tau_\star$ starts from $\approx 2.2\tau_\eta$ at $\hat{\St}=0$  \citep[as documented for tracers in][]{yeung1989lagrangian}, and then increases monotonically with the isotropically-averaged Stokes number.

\section{Concluding remarks}
\label{sec:conclusions}
In this study, we investigated the dynamics of heavy spheroidal particles transported by a homogeneous isotropic turbulent flow. Our numerical simulations provided strong evidence that the isotropically-averaged Stokes number, as introduced by \citet{shapiro1993deposition}, and based on the effective response time obtained by averaging the particle mobility tensor over all possible orientations, fully captures the shape dependence in their translational dynamics. Whether this definition is more relevant than the one proposed by \citet{fan1995dispersion}, where the drag tensor is averaged instead, remains to be determined by future work with much more precise statistics. Nevertheless, our results demonstrate that the single-particle statistics of an inertial spheroid, including the probability distribution of velocity and acceleration and their time correlations, are similar to those of an equivalent spherical particle whose effective diameter depends on its aspect ratio, opening new ways to develop macroscopic models for the turbulent transport of non-spherical particles. Our observations suggest that angular and translational dynamics are largely uncorrelated, which can be explained by the fast timescales associated with particle spinning and tumbling. Specifically, the response to fluid-flow rotation is at least twice as fast as translational equilibration.

While our study focused on statistically isotropic flows, it is worth noting that intricate relations between rotation and translation could arise in anisotropic situations, such as in flows with a mean shear or in the presence of boundaries. In these cases, prolate particles tend to orient in the direction of the flow, whereas oblate particles are more likely to align in the direction of its gradient. Turbulent structures in such flows also exhibit strong anisotropies, and the response of spheroids to turbulent fluctuations may depend non-trivially on their shape. Recent experimental results by \citet{baker2022experimental} suggest that near to the walls of a channel flow, rods tumble more frequently than disks, and the latter respond more slowly to fluid-velocity fluctuations. These observations imply that efficient macroscopic models for the transport of spheroids by anisotropic flow may need to weigh differently the components of the particle mobility tensor. Further work is needed to develop a comprehensive understanding of the effects of anisotropies on the turbulent transport of non-spherical particles.

Finally, many questions remain unanswered regarding two-particle statistics of spheroids in turbulence. While we observed a decoupling between translation and rotation at the single-particle level, this may not hold true for their relative motion. Previous studies have shown that, in the absence of inertia, spheroidal particles tend to align with the eigendirections of the Cauchy-Green tensor \citep{ni2014alignment}, and we expect these correlations to persist for small-inertia particles. In this case, particles concentrate on dynamically-evolving attractors with a fractal structure related to the stretching and compression directions of the Cauchy--Green tensor. This could lead to intricate relationships between clustering and alignment, possibly making it impossible to describe particle spatial patterns in terms of a single shape-dependent Stokes number. For particles with large inertia, the presence of caustics, where particles with very different histories come arbitrarily close to each other, competes with fractal clustering. Such particles are likely to have significantly different orientations, making it even more challenging to predict how their relative motion depends on their shape.

\backsection[Acknowledgements]{We are deeply grateful to C.~Siewert for numerous discussions and we acknowledge the precious help of H.~Homann and J.-I.~Polanco for the implementation of spheroids in the \textit{LaTu} code.}

\backsection[Funding]{Computational resources were provided by GENCI (grant IDRIS 2019-A0062A10800) and by the OPAL infrastructure from Universit\'e C\^ote d'Azur. This work received support from the UCA-JEDI Future Investments, funded by the French government (grant no. ANR-15-IDEX-01), and from the Agence Nationale de la Recherche (grant no. ANR-21-CE30-0040-01).}

\bibliographystyle{jfm}

\begin{thebibliography}{39}
\expandafter\ifx\csname natexlab\endcsname\relax\def\natexlab#1{#1}\fi
\def\au#1{#1} \def\ed#1{#1} \def\yr#1{#1}\def\at#1{#1}\def\jt#1{\textit{#1}}
  \def\bt#1{#1}\def\bvol#1{\textbf{#1}} \def\vol#1{#1} \def\pg#1{#1}
  \def\publ#1{#1}\def\arxiv#1{#1}\def\org#1{#1}\def\st#1{\textit{#1}}

\bibitem[Anand {\em et~al.\/}(2020)Anand, Ray \&
  Subramanian]{anand2020orientation}
{\sc \au{Anand, P.}, \au{Ray, S.S.} \& \au{Subramanian, G.}} \yr{2020}
  \at{Orientation dynamics of sedimenting anisotropic particles in turbulence}.
   \jt{Phys. Rev. Lett.}  \bvol{125},  \pg{034501}.

\bibitem[Baker \& Coletti(2022)]{baker2022experimental}
{\sc \au{Baker, L.J.} \& \au{Coletti, F.}} \yr{2022}  \at{Experimental
  investigation of inertial fibres and disks in a turbulent boundary layer}.
  \jt{J. Fluid Mech.}  \bvol{943},  \pg{A27}.

\bibitem[Bec {\em et~al.\/}(2006)Bec, Biferale, Boffetta, Celani, Cencini,
  Lanotte, Musacchio \& Toschi]{bec2005acceleration}
{\sc \au{Bec, J.}, \au{Biferale, L.}, \au{Boffetta, G.}, \au{Celani, A.},
  \au{Cencini, M.}, \au{Lanotte, A.}, \au{Musacchio, S.} \& \au{Toschi, F.}}
  \yr{2006}  \at{Acceleration statistics of heavy particles in turbulence}.
  \jt{J. Fluid Mech.}  \bvol{550},  \pg{349--358}.

\bibitem[Brandt \& Coletti(2022)]{brandt2022particle}
{\sc \au{Brandt, L.} \& \au{Coletti, F.}} \yr{2022}  \at{Particle-laden
  turbulence: progress and perspectives}.  \jt{Annu. Rev. Fluid Mech.}
  \bvol{54},  \pg{159--189}.

\bibitem[Brenner(1964)]{brenner1964stokes}
{\sc \au{Brenner, H.}} \yr{1964}  \at{The {S}tokes resistance of a slightly
  deformed sphere}.  \jt{Chem. Eng. Sci.}  \bvol{19},  \pg{519--539}.

\bibitem[Byron {\em et~al.\/}(2015)Byron, Einarsson, Gustavsson, Voth, Mehlig
  \& Variano]{byron2015shape}
{\sc \au{Byron, M.}, \au{Einarsson, J.}, \au{Gustavsson, K.}, \au{Voth, G.A.},
  \au{Mehlig, B.} \& \au{Variano, E.}} \yr{2015}  \at{Shape-dependence of
  particle rotation in isotropic turbulence}.  \jt{Phys. Fluids}  \bvol{27},
  \pg{035101}.

\bibitem[Challabotla {\em et~al.\/}(2015)Challabotla, Zhao \&
  Andersson]{challabotla2015orientation}
{\sc \au{Challabotla, N.}, \au{Zhao, L.} \& \au{Andersson, H.}} \yr{2015}
  \at{Orientation and rotation of inertial disk particles in wall turbulence}.
  \jt{J. Fluid Mech.}  \bvol{766}.

\bibitem[Del~Bello {\em et~al.\/}(2015)Del~Bello, Taddeucci, Scarlato,
  Giacalone \& Cesaroni]{del2015experimental}
{\sc \au{Del~Bello, E.}, \au{Taddeucci, J.}, \au{Scarlato, P.}, \au{Giacalone,
  E.} \& \au{Cesaroni, C.}} \yr{2015}  \at{Experimental investigation of the
  aggregation-disaggregation of colliding volcanic ash particles in turbulent,
  low-humidity suspensions}.  \jt{Geophys. Res. Lett.}  \bvol{42},
  \pg{1068--1075}.

\bibitem[Fan \& Ahmadi(1995)]{fan1995dispersion}
{\sc \au{Fan, F.-G.} \& \au{Ahmadi, G.}} \yr{1995}  \at{Dispersion of
  ellipsoidal particles in an isotropic pseudo-turbulent flow field}.  \jt{ASME
  J.\ Fluids Eng.}  \bvol{117},  \pg{154--161}.

\bibitem[Grythe {\em et~al.\/}(2014)Grythe, Str{\"o}m, Krejci, Quinn \&
  Stohl]{grythe2014review}
{\sc \au{Grythe, H.}, \au{Str{\"o}m, J.}, \au{Krejci, R.}, \au{Quinn, P.} \&
  \au{Stohl, A.}} \yr{2014}  \at{A review of sea-spray aerosol source functions
  using a large global set of sea salt aerosol concentration measurements}.
  \jt{Atmospheric Chem. Phys.}  \bvol{14},  \pg{1277--1297}.

\bibitem[Gustavsson {\em et~al.\/}(2014)Gustavsson, Einarsson \&
  Mehlig]{gustavsson2014tumbling}
{\sc \au{Gustavsson, K.}, \au{Einarsson, J.} \& \au{Mehlig, B.}} \yr{2014}
  \at{Tumbling of small axisymmetric particles in random and turbulent flows}.
  \jt{Phys. Rev. Lett.}  \bvol{112},  \pg{014501}.

\bibitem[Gustavsson {\em et~al.\/}(2019)Gustavsson, Sheikh, Lopez, Naso, Pumir
  \& Mehlig]{gustavsson2019effect}
{\sc \au{Gustavsson, K.}, \au{Sheikh, M.Z.}, \au{Lopez, D.}, \au{Naso, A.},
  \au{Pumir, A.} \& \au{Mehlig, B}} \yr{2019}  \at{Effect of fluid inertia on
  the orientation of a small prolate spheroid settling in turbulence}.  \jt{New
  J. Phys.}  \bvol{21}~(8),  \pg{083008}.

\bibitem[Homann {\em et~al.\/}(2007)Homann, Dreher \& Grauer]{homann2007impact}
{\sc \au{Homann, H.}, \au{Dreher, J.} \& \au{Grauer, R.}} \yr{2007}  \at{Impact
  of the floating-point precision and interpolation scheme on the results of
  dns of turbulence by pseudo-spectral codes}.  \jt{Comput. Phys. Comm.}
  \bvol{177}~(7),  \pg{560--565}.

\bibitem[Horowitz {\em et~al.\/}(2020)Horowitz, Holmes, Wright, Sherwen, Wang,
  Evans, Huang, Jaegl{\'e}, Chen, Zhai \& Alexander]{horowitz2020effects}
{\sc \au{Horowitz, H.M.}, \au{Holmes, C.}, \au{Wright, A.}, \au{Sherwen, T.},
  \au{Wang, X.}, \au{Evans, M.}, \au{Huang, J.}, \au{Jaegl{\'e}, L.}, \au{Chen,
  Q.}, \au{Zhai, S.} \& \au{Alexander, B.}} \yr{2020}  \at{Effects of sea salt
  aerosol emissions for marine cloud brightening on atmospheric chemistry:
  {I}mplications for radiative forcing}.  \jt{Geophys. Res. Lett.}  \bvol{47},
  \pg{e2019GL085838}.

\bibitem[Jeffery(1922)]{jeffery1922motion}
{\sc \au{Jeffery, G.}} \yr{1922}  \at{The motion of ellipsoidal particles
  immersed in a viscous fluid}.  \jt{Proc. R. Soc. A}  \bvol{102}~(715),
  \pg{161--179}.

\bibitem[Jucha {\em et~al.\/}(2018)Jucha, Naso, L{\'e}v{\^e}que \&
  Pumir]{jucha2018settling}
{\sc \au{Jucha, J.}, \au{Naso, A.}, \au{L{\'e}v{\^e}que, E.} \& \au{Pumir, A.}}
  \yr{2018}  \at{Settling and collision between small ice crystals in turbulent
  flows}.  \jt{Phys. Rev. Fluids}  \bvol{3}~(1),  \pg{014604}.

\bibitem[Klett(1995)]{klett1995orientation}
{\sc \au{Klett, James~D}} \yr{1995}  \at{Orientation model for particles in
  turbulence}.  \jt{J. Atmos. Sci.}  \bvol{52}~(12),  \pg{2276--2285}.

\bibitem[Leblanc {\em et~al.\/}(2018)Leblanc, Queguiner, Diaz, Cornet,
  Michel-Rodriguez, Durrieu~de Madron, Bowler, Malviya, Thyssen, Gr{\'e}gori
  {\em et~al.\/}]{leblanc2018nanoplanktonic}
{\sc \au{Leblanc, K.}, \au{Queguiner, B.}, \au{Diaz, F.}, \au{Cornet, V.},
  \au{Michel-Rodriguez, M.}, \au{Durrieu~de Madron, X.}, \au{Bowler, C.},
  \au{Malviya, S.}, \au{Thyssen, M.}, \au{Gr{\'e}gori, G.} \& \au{others}}
  \yr{2018}  \at{Nanoplanktonic diatoms are globally overlooked but play a role
  in spring blooms and carbon export}.  \jt{Nat. Commun.}  \bvol{9},  \pg{953}.

\bibitem[Marchioli \& Soldati(2013)]{marchioli2013rotation}
{\sc \au{Marchioli, C.} \& \au{Soldati, A.}} \yr{2013}  \at{Rotation statistics
  of fibers in wall shear turbulence}.  \jt{Acta Mech.}  \bvol{224},
  \pg{2311--2329}.

\bibitem[Marchioli {\em et~al.\/}(2016)Marchioli, Zhao \&
  Andersson]{marchioli2016relative}
{\sc \au{Marchioli, C.}, \au{Zhao, L.} \& \au{Andersson, H.I.}} \yr{2016}
  \at{On the relative rotational motion between rigid fibers and fluid in
  turbulent channel flow}.  \jt{Phys. Fluids}  \bvol{28},  \pg{013301}.

\bibitem[Marcus {\em et~al.\/}(2014)Marcus, Parsa, Kramel, Ni \&
  Voth]{marcus2014measurements}
{\sc \au{Marcus, G.G.}, \au{Parsa, S.}, \au{Kramel, S.}, \au{Ni, R.} \&
  \au{Voth, G.A.}} \yr{2014}  \at{Measurements of the solid-body rotation of
  anisotropic particles in 3{D} turbulence}.  \jt{New J. Phys.}  \bvol{16},
  \pg{102001}.

\bibitem[Mortensen {\em et~al.\/}(2008)Mortensen, Andersson, Gillissen \&
  Boersma]{mortensen2008dynamics}
{\sc \au{Mortensen, PH.}, \au{Andersson, HI.}, \au{Gillissen, JJJ.} \&
  \au{Boersma, BJ.}} \yr{2008}  \at{Dynamics of prolate ellipsoidal particles
  in a turbulent channel flow}.  \jt{Phys. Fluids}  \bvol{20},  \pg{093302}.

\bibitem[Ni {\em et~al.\/}(2014)Ni, Ouellette \& Voth]{ni2014alignment}
{\sc \au{Ni, R.}, \au{Ouellette, N.T.} \& \au{Voth, G.A.}} \yr{2014}
  \at{Alignment of vorticity and rods with {L}agrangian fluid stretching in
  turbulence}.  \jt{J. Fluid Mech.}  \bvol{743},  \pg{R3}.

\bibitem[Parsa {\em et~al.\/}(2012)Parsa, Calzavarini, Toschi \&
  Voth]{parsa2012rotation}
{\sc \au{Parsa, S.}, \au{Calzavarini, E.}, \au{Toschi, F.} \& \au{Voth, G.A}}
  \yr{2012}  \at{Rotation rate of rods in turbulent fluid flow}.  \jt{Phys.
  Rev. Lett.}  \bvol{109},  \pg{134501}.

\bibitem[Prata \& Lynch(2019)]{prata2019passive}
{\sc \au{Prata, F.} \& \au{Lynch, M.}} \yr{2019}  \at{Passive earth
  observations of volcanic clouds in the atmosphere}.  \jt{Atmosphere}
  \bvol{10}~(4),  \pg{199}.

\bibitem[Pumir \& Wilkinson(2011)]{pumir2011orientation}
{\sc \au{Pumir, A.} \& \au{Wilkinson, M.}} \yr{2011}  \at{Orientation
  statistics of small particles in turbulence}.  \jt{New J. Phys.}  \bvol{13},
  \pg{093030}.

\bibitem[Roy {\em et~al.\/}(2018)Roy, Gupta \& Ray]{roy2018inertial}
{\sc \au{Roy, A.}, \au{Gupta, A.} \& \au{Ray, S.S.}} \yr{2018}  \at{Inertial
  spheroids in homogeneous, isotropic turbulence}.  \jt{Phys. Rev. E}
  \bvol{98},  \pg{021101}.

\bibitem[Salazar \& Collins(2012)]{salazar2012inertial_vel}
{\sc \au{Salazar, J.P.L.C.} \& \au{Collins, L.R.}} \yr{2012}  \at{Inertial
  particle relative velocity statistics in homogeneous isotropic turbulence}.
  \jt{J. Fluid Mech.}  \bvol{696},  \pg{45--66}.

\bibitem[Sengupta {\em et~al.\/}(2017)Sengupta, Carrara \&
  Stocker]{sengupta2017phytoplankton}
{\sc \au{Sengupta, A.}, \au{Carrara, F.} \& \au{Stocker, R.}} \yr{2017}
  \at{Phytoplankton can actively diversify their migration strategy in response
  to turbulent cues}.  \jt{Nature}  \bvol{543}~(7646),  \pg{555--558}.

\bibitem[Shapiro \& Goldenberg(1993)]{shapiro1993deposition}
{\sc \au{Shapiro, M.} \& \au{Goldenberg, M.}} \yr{1993}  \at{Deposition of
  glass fiber particles from turbulent air flow in a pipe}.  \jt{J. Aerosol
  Sci.}  \bvol{24},  \pg{65--87}.

\bibitem[Sheikh {\em et~al.\/}(2020)Sheikh, Gustavsson, Lopez, L{\'e}v{\^e}que,
  Mehlig, Pumir \& Naso]{sheikh2020importance}
{\sc \au{Sheikh, M.Z.}, \au{Gustavsson, K.}, \au{Lopez, D.},
  \au{L{\'e}v{\^e}que, E.}, \au{Mehlig, B.}, \au{Pumir, A.} \& \au{Naso, A.}}
  \yr{2020}  \at{Importance of fluid inertia for the orientation of spheroids
  settling in turbulent flow}.  \jt{J. Fluid Mech.}  \bvol{886},  \pg{A9}.

\bibitem[Siewert {\em et~al.\/}(2014{\natexlab{{\em a\/}}})Siewert, Kunnen,
  Meinke \& Schr{\"o}der]{siewert2014orientation}
{\sc \au{Siewert, C.}, \au{Kunnen, R.P.J.}, \au{Meinke, M.} \&
  \au{Schr{\"o}der, W.}} \yr{2014{\natexlab{{\em a\/}}}}  \at{Orientation
  statistics and settling velocity of ellipsoids in decaying turbulence}.
  \jt{Atmos. Res.}  \bvol{142},  \pg{45--56}.

\bibitem[Siewert {\em et~al.\/}(2014{\natexlab{{\em b\/}}})Siewert, Kunnen \&
  Schr{\"o}der]{siewert2014collision}
{\sc \au{Siewert, C.}, \au{Kunnen, R.P.J.} \& \au{Schr{\"o}der, W.}}
  \yr{2014{\natexlab{{\em b\/}}}}  \at{Collision rates of small ellipsoids
  settling in turbulence}.  \jt{J. Fluid Mech.}  \bvol{758},  \pg{686--701}.

\bibitem[Toschi \& Bodenschatz(2009)]{toschi2009lagrangian}
{\sc \au{Toschi, F.} \& \au{Bodenschatz, E.}} \yr{2009}  \at{Lagrangian
  properties of particles in turbulence}.  \jt{Annu. Rev. Fluid Mech.}
  \bvol{41},  \pg{375--404}.

\bibitem[Voth \& Soldati(2017)]{voth2017anisotropic}
{\sc \au{Voth, G.A.} \& \au{Soldati, A.}} \yr{2017}  \at{Anisotropic particles
  in turbulence}.  \jt{Annu. Rev. Fluid Mech.}  \bvol{49},  \pg{249--276}.

\bibitem[Yeung \& Pope(1989)]{yeung1989lagrangian}
{\sc \au{Yeung, P.-K.} \& \au{Pope, S.}} \yr{1989}  \at{Lagrangian statistics
  from direct numerical simulations of isotropic turbulence}.  \jt{J. Fluid
  Mech.}  \bvol{207},  \pg{531--586}.

\bibitem[Yeung {\em et~al.\/}(2006)Yeung, Pope, Lamorgese \&
  Donzis]{yeung2006acceleration}
{\sc \au{Yeung, P.-K.}, \au{Pope, S.}, \au{Lamorgese, A.} \& \au{Donzis, D.}}
  \yr{2006}  \at{Acceleration and dissipation statistics of numerically
  simulated isotropic turbulence}.  \jt{Phys. Fluids}  \bvol{18}~(6),
  \pg{065103}.

\bibitem[Zhang {\em et~al.\/}(2001)Zhang, Ahmadi, Fan \&
  McLaughlin]{zhang2001ellipsoidal}
{\sc \au{Zhang, H.}, \au{Ahmadi, G.}, \au{Fan, F.G.} \& \au{McLaughlin, J.B.}}
  \yr{2001}  \at{Ellipsoidal particles transport and deposition in turbulent
  channel flows}.  \jt{Int. J. Multiph. Flow}  \bvol{27}~(6),  \pg{971--1009}.

\bibitem[Zhao {\em et~al.\/}(2015)Zhao, Challabotla, Andersson \&
  Variano]{zhao2015rotation}
{\sc \au{Zhao, L.}, \au{Challabotla, N.}, \au{Andersson, H.} \& \au{Variano,
  E.}} \yr{2015}  \at{Rotation of nonspherical particles in turbulent channel
  flow}.  \jt{Phys. Rev. lett.}  \bvol{115}~(24),  \pg{244501}.

\end{thebibliography}

\end{document}